%% file: main.tex
\documentclass{article}
\usepackage{spconf,amsmath,graphicx}
\usepackage{booktabs}
\usepackage{multirow}
\usepackage[font=small]{caption}
\usepackage{bm}
\usepackage{bbm}
\usepackage{tabularx}
\usepackage[normalem]{ulem}
\usepackage[table]{xcolor} 
\usepackage{xspace}
\usepackage{enumitem}
\usepackage{arydshln}
\usepackage{array}
\usepackage{lipsum}

\useunder{\uline}{\ul}{}

\usepackage[pagebackref,breaklinks,colorlinks]{hyperref}

\usepackage[capitalize]{cleveref}
\crefname{section}{Sec.}{Secs.}
\Crefname{section}{Section}{Sections}
\Crefname{table}{Table}{Tables}
\crefname{table}{Tab.}{Tabs.}

\usepackage{enumitem}
\setlist{nosep, leftmargin=14pt}

\usepackage{mwe} 

\def\etal{\emph{et al}.}
\def\eg{\emph{e}.\emph{g}.}
\def\ie{\emph{i}.\emph{e}.}

\newcommand{\mysection}[1]{\vspace{-0.5mm}\section{#1}\vspace{-0.5mm}}
\newcommand{\mysubsection}[1]{\vspace{-0.5mm}\subsection{#1}\vspace{-0.5mm}}

\newcommand\blfootnote[1]{%
  \begingroup
  \renewcommand\thefootnote{}\footnote{#1}%
  \addtocounter{footnote}{-1}%
  \endgroup
}

\title{SIFT-DBT: Self-supervised Initialization and Fine-Tuning for Imbalanced Digital Breast Tomosynthesis Image Classification}

\name{Yuexi Du$^{\star}$ \quad Regina J. Hooley$^{\dagger}$ \quad John Lewin$^{\dagger}$ \quad Nicha C. Dvornek$^{\star\dagger}$}
\address{$^{\star}$ Department of Biomedical Engineering, Yale University, New Haven, CT\\
$^{\dagger}$ Department of Radiology \& Biomedical Imaging, Yale School of Medicine, New Haven, CT}

\begin{document}
%
\maketitle
\input{parts/0_abstract}
\blfootnote{\textsuperscript{\textcopyright}2024 IEEE.  Personal use of this material is permitted.  Permission from IEEE must be obtained for all other uses, in any current or future media, including reprinting/republishing this material for advertising or promotional purposes, creating new collective works, for resale or redistribution to servers or lists, or reuse of any copyrighted component of this work in other works.}

\input{parts/1_intro}

\input{parts/2_method}

\input{parts/3_experiment}

\input{parts/4_conclusion}

\input{parts/5_acknowledge}

\bibliographystyle{IEEEbib}
\bibliography{strings,refs}

\end{document}

%% file: parts/0_abstract.tex
\begin{abstract}
    Digital Breast Tomosynthesis (DBT) is a widely used medical imaging modality for breast cancer screening and diagnosis, offering higher spatial resolution and greater detail through its 3D-like breast volume imaging capability. However, the increased data volume also introduces pronounced data imbalance challenges, where only a small fraction of the volume contains suspicious tissue. This further exacerbates the data imbalance due to the case-level distribution in real-world data and leads to learning a trivial classification model that only predicts the majority class. To address this, we propose a novel method using view-level contrastive \textbf{S}elf-supervised \textbf{I}nitialization and \textbf{F}ine-\textbf{T}uning for identifying abnormal \textbf{DBT} images, namely \textbf{SIFT-DBT}.
    We further introduce a patch-level multi-instance learning method to preserve spatial resolution. The proposed method achieves 92.69\% volume-wise AUC on an evaluation of 970 unique studies. \footnote{Code is available at \url{https://github.com/XYPB/SIFT_DBT}.}
\end{abstract}

%% file: parts/1_intro.tex
\section{Introduction}
\vspace{-0.8mm}

Breast cancer is the most commonly diagnosed cancer globally and the leading cause of cancer-related mortality in women \cite{GLOBOCAN}. Timely and accurate radiologic imaging plays a pivotal role in the early detection of potential lesions, proving to be a crucial component in the effective management of breast cancer. Over the past 10 years, Digital Breast Tomosynthesis (DBT) has emerged as a powerful imaging tool in breast cancer detection \cite{hooley2017advances}.
Different from traditional 2D mammography, DBT provides a more comprehensive z-dimensional view of breast tissue by rotating the X-ray generator around the breast to reconstruct the 3D information,
resulting in significantly enhanced resolution of tissue details. Meanwhile, DBT also provides two-view information (\ie, craniocaudal (CC) and mediolateral oblique (MLO)) for each side of each study. However, the increase in imaging information also amplifies the challenge of data imbalance. In the currently largest public DBT dataset BCS-DBT~\cite{konz2023competition, buda2020detection}, there are only 101 studies (comprising 224 volumes) labeled as abnormal among all annotated 4,838 studies (encompassing 19,148 volumes). Despite a lesion appearing in multiple slices within each abnormal volume (containing 62 slices on average), only one single slice is annotated with a bounding box of $\sim$2\% of the area. Thus, the data imbalance between normal and abnormal tissue becomes even more evident. Such extreme imbalance can lead to the failure of various modern computer-aided DBT diagnosis methods. Moreover, the increasing amount of data and its imbalanced distribution significantly intensifies radiologists' workload, with most of the time spent on filtering out healthy scans.

While supervised deep learning methods have shown promising performance in medical image analysis, the challenge of imbalanced data remains a major issue~\cite{zhang2023deep, li2023hierarchical}. Numerous strategies have been devised to mitigate this, \eg, over- or under-sampling, and focal loss~\cite{lin2017focal}. However, these methods still fall short when faced with extreme data imbalance, as they were solely trained to fit the imbalanced labels, neglecting the potential semantic information within the image.
Meanwhile, prior DBT studies~\cite{lee2023transformer, tardy2021trainable} curate balanced data, which do not reflect the real-world distribution.

To address this shortcoming, we propose a novel approach, called SIFT-DBT, for DBT image classification employing self-supervised contrastive learning that exploits the information within the data itself. The contrastive learning paradigm encourages the model to focus on structural and semantic information rather than features that are only related to class distribution. We further propose a local multi-patch fine-tuning method, striking a balance between image resolution and computational efficiency while enhancing slice-level classification performance. The slice-wise scores are further aggregated to obtain volume-level predictions. Our SIFT-DBT outperforms multiple baselines on a test subset of 970 studies from the public BCS-DBT~\cite{buda2020detection} dataset.
The success of our method has the potential to help radiologists quickly identify higher-risk scans and substantially improve the efficiency of the screening process by filtering out normal exams.

%% file: parts/2_method.tex
\input{floats/fig_method}
\mysection{Method}

Our SIFT-DBT framework consists of a contrastive learning paradigm with positive-pair selection defined specifically for DBT studies, followed by a local multi-patch fine-tuning strategy for abnormal slice and volume detection (\cref{fig:method}).

\mysubsection{Contrastive Learning Paradigm Review}

Given an input $x$, contrastive learning minimizes the similarity between $x$ and its positive sample $x_+$ in the embedding space while maximizing the similarity with respect to its negative sample $x_-$. Suppose there are $S$ samples in memory, including one positive and $S-1$ negative. The optimization objective, \ie, InfoNCE loss~\cite{oord2018representation}, can be written as 
\vspace{-1mm}
\begin{equation}
\vspace{-1mm}
\mathcal{L}_{InfoNCE} = -\log\frac{\exp(x\cdot x_+/\tau)}{\sum^S_{i=1}\exp(x\cdot x_i/\tau)},
\end{equation}
where $\tau$ is the temperature constant. Here, we note that the similarity is computed in the embedding space encoded with two encoders $f_\theta$ and $g_{\theta'}$ with the same architecture, and $g_{\theta'}$ is the momentum encoder with its parameter updated by exponential moving average (EMA) and gradient stopped, \ie, 
\vspace{-2mm}
\begin{equation}
\vspace{-1mm}
\theta'\leftarrow m\theta' + (1-m)\theta,
\end{equation} 
where $m$ is the momentum of the EMA update. This objective will naturally encourage the model to learn the structural and semantic similarity between positive pairs. We follow MoCo v2~\cite{he2020momentum} for the pre-training architecture, where there is a memory bank to store more negative samples to improve the quality of negative pairs (\cref{fig:method}(a)). Our method could also be similarly integrated with other contrastive learning methods.

\mysubsection{DBT Metadata Based Positive Pair Selection}

One of the most critical aspects of successful self-supervised contrastive learning is how to build robust and reasonable positive pairs without labels. The traditional contrastive learning method uses aggressive data augmentation to construct a positive pair from the same image. However, this is not sufficient in DBT data, where the internal variance of data within the same image is relatively small, and the global similarity is larger than in natural images. Almost all of the DBT images contain a similar orientation and tissue including mammary glands, ducts, and fatty tissue.

Inspired by Vu~\etal~\cite{vu2021medaug}, we use the DBT metadata to build a more robust positive pair. Apart from the augmented pairs from the same slice (\ie, image), we also treat two slices from the same volume and from different views of the breast as positive pairs to each other. It is natural to treat different slices from the same volume as positive since they are obtained from the same breast but at different depths. Similar reasoning holds for the two different views of the same side, as they are the same breast scanned from a different angle (CC or MLO). However, different from Vu~\etal, we keep slices from the different sides and other slices from the same patient as \textit{negative} samples. This is because the potential malignant tumor may appear on only the left or right breast. Meanwhile, the other studies from the same patient may come from different time points and may have different characteristics according to the development of potential tumors.

We adopt the same data augmentation as MoCo~\cite{he2020momentum} but reduce the resize crop and color jittering strength to maintain a similar level of tissue structure and overall intensity. Additionally, we randomly sample $x_+$ from a slice in the other view with $50\%$ probability, otherwise, $x_+$ is sampled from a neighboring slice within a range of $k$ with respect to $x$.

\mysubsection{Local Multi-Patch Fine-tuning}

Although the pre-trained model with contrastive learning already provides robust and identifiable features, making use of these features directly does not provide optimal performance. Furthermore, considering the need to maintain the image resolution and computational cost to run the model on the original size, fine-tuning with a full slice becomes impractical.

We introduce the local multi-patch fine-tuning method to address these issues. During training, we randomly sample a sub-patch and fine-tune the model in a discriminative manner~\cite{howard2018universal}. Namely, we decrease the learning rate by a factor $\eta=2.8$ for each block of the model. This helps maintain the pre-trained low-level feature extractor which is already robust enough and focuses on tuning the deeper layers that contain more semantic information. At test time, we sample $N$ sub-patches randomly from the input slice and aggregate the predictions $p_i$ with the function $f_a=avg(\cdot)$ to obtain the final prediction $p$ (\cref{fig:method}(b)).

Additionally, we restrict that the patch sampled from abnormal slices during training time must contain the region of the annotated tumor. This can make full use of the annotation and force the model to focus on the suspicious region. We do not have this restriction during test time. Lastly, following the common training practice for imbalanced data distribution, we apply over-sampling on the minority class and under-sampling on the majority class to ensure each batch is diverse and balanced.

\input{floats/tab_slice}
\input{floats/fig_slice_auc}

\vspace{-2mm}
\mysubsection{Volume-Level Classification from Patch Prediction}

While our training method is designed for slice-level classification, we further extend to volume-level classification with parameter-free prediction based on slice-level results. We aggregate slice-level prediction scores from the same volume by taking the \textit{maximum} abnormal score from all slices and treating it as the volume-level prediction score. Then, we select the volume-level score threshold that minimizes the gap between normal and abnormal recall, which best balances the performance for both classes under an imbalanced situation.

%% file: floats/fig_method.tex
\begin{figure*}[!t]
    \centering
    \includegraphics[width=0.96\textwidth]{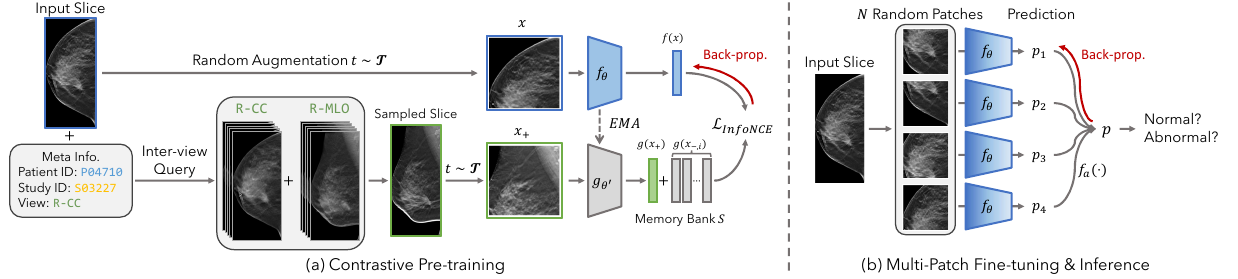}
    \vspace{-3mm}
    \caption{{\bf Proposed SIFT-DBT framework}.
    (a) We use random augmentation $t\sim\mathcal{T}$ with inter-view and inter-slice sampling to get the positive pair inputs for pre-training. (b) During fine-tuning, we updated the model with only one sub-patch sampled from the full image. The blue gradient color within the model indicates we use the discriminative learning rate for each block.}  
    \label{fig:method}
    \vspace{-3mm}
\end{figure*}  

%% file: floats/tab_slice.tex
\begin{table}[!t]
\centering
\resizebox{\columnwidth}{!}
{

\begin{tabular}{@{}llcccccc@{}}
\specialrule{.1em}{.05em}{.05em}
\multicolumn{1}{c}{\multirow{2}{*}{Method}} & \multicolumn{1}{c}{\multirow{2}{*}{Model}} & \multicolumn{6}{c}{\textbf{BCS-DBT Slice-level Results}} \\ \cmidrule(l){3-8} 
\multicolumn{1}{c}{} & \multicolumn{1}{c}{} & AUC & NPV & NR & AR & SP@87 & SP@80 \\ \specialrule{.08em}{.05em}{.05em}
\multirow{2}{*}{Rand. Init.} & CNN & 87.94 & 99.91 & 59.10 & \textbf{90.31} & 71.33 & 77.23 \\
 & ResNet50 & 85.03 & 99.76 & 86.03 & 61.25 & 64.22 & 68.85 \\ \midrule

MoCo v2 & \multirow{3}{*}{ResNet50} & 89.80 & 99.75 & {\ul 93.40} & 56.56 & 76.65 & 83.74 \\
MedAug &  & 86.89 & 99.86 & 78.97 & 79.37 & 69.80 & 78.68 \\
Inter-slice &  & 88.67 & 99.84 & 87.42 & 74.69 & 77.29 & 84.38 \\ \midrule
\multirow{2}{*}{Ours\textsubscript{$N$=1}} & CNN & 89.73 & 99.83 & 86.36 & 73.44 & 76.78 & 83.62 \\
 & ResNet50 & 90.31 & 99.89 & 81.13 & 83.13 & 76.69 & 83.05 \\\hdashline 
\multirow{2}{*}{Ours} & CNN & \textbf{94.26} & {\ul 99.95} & \textbf{98.98} & 60.94 & \textbf{86.85} & {\ul 92.98} \\
 & ResNet50 & {\ul 93.67} & \textbf{99.98} & 76.38 & {\ul 88.44} & {\ul 82.20} & \textbf{97.16} \\ \specialrule{.1em}{.05em}{.05em}
\end{tabular}
}
\vspace{-2mm}
\caption{\textbf{Slice-level Classification Results.} 
We highlight the top result in bold and the second-best with an underline. NR, Normal Recall; AR, Abnormal Recall; SP@\#, Specificity at \#\% sensitivity.} 
\label{tab:slice}
\vspace{-3mm}
\end{table}

%% file: floats/fig_slice_auc.tex
\begin{figure}[!t]
    \centering
    \includegraphics[width=1.0\columnwidth]{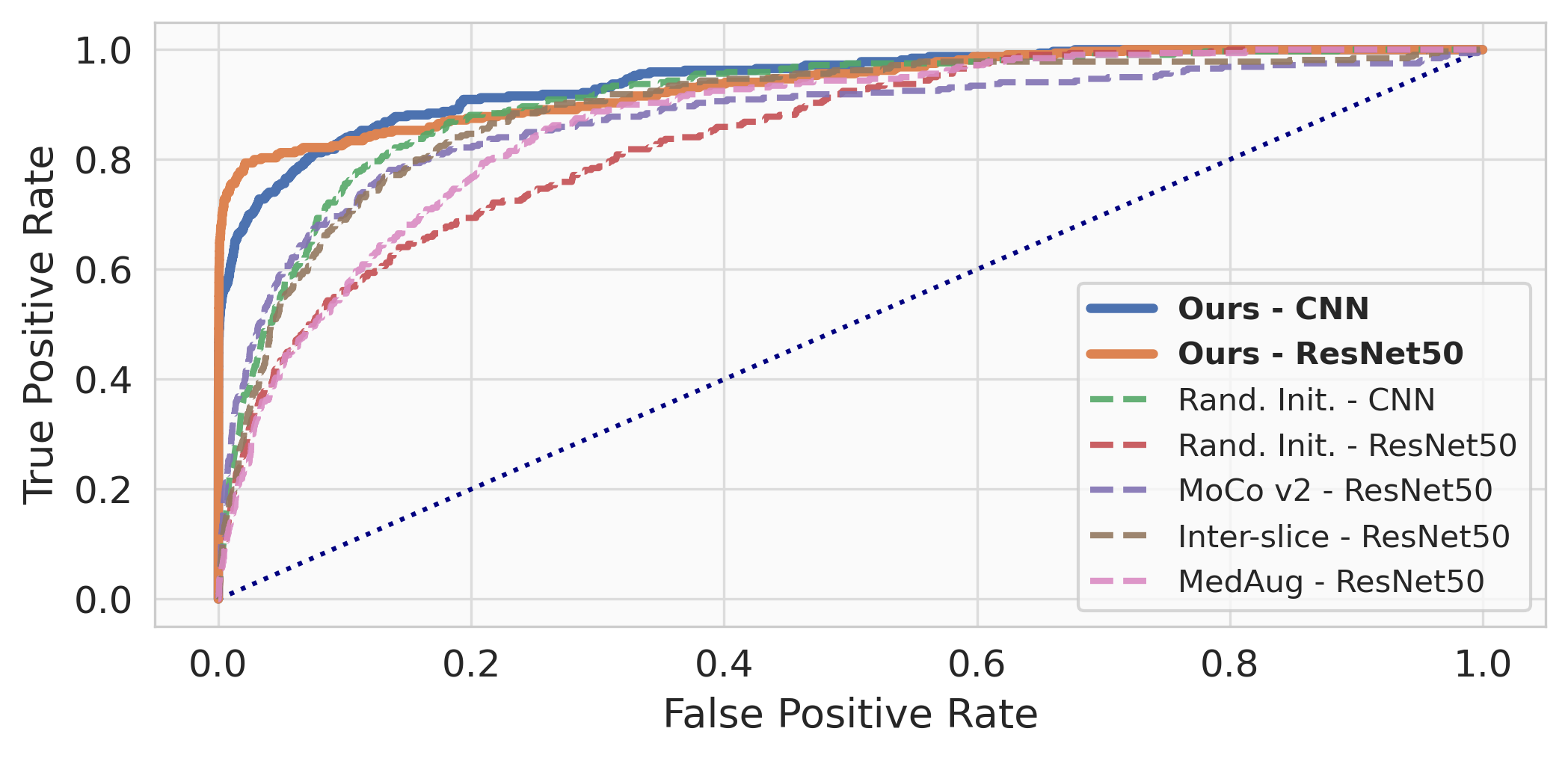}
    \vspace{-7mm}
    \caption{{\bf Slice-level ROC curve}. We plot the slice-level ROC curve for the baselines. Our method is plotted with a solid line.}
    \label{fig:slice_roc}
    \vspace{-4mm}
\end{figure}  

%% file: parts/3_experiment.tex
\mysection{Experiments}

\mysubsection{Experiment Setup}

\textbf{Data.} We use the BCS-DBT~\cite{buda2020detection} dataset, which was collected by Duke University and passed ethical evaluation. Since only the annotation for the training set of the BCS-DBT~\cite{buda2020detection} dataset is available publicly, we split the training set subject-wise into training, validation, and test sets with a ratio of 7:1:2. We further dropped the ``Actionable'' class from the original dataset since the definition is ambiguous~\cite{buda2020detection}. We then treat both the benign and cancer groups as abnormal. As a result, the training, validation, and test sets contain 12,819 volumes with 139 abnormal volumes, 1,838 volumes with 22 abnormal volumes, and 3,775 volumes with 39 abnormal volumes, respectively. We resize all the slices to 1024 pixels on the short side and then apply Otsu thresholding to remove the extra background regions across the same volume. 

\input{floats/tab_fintune}

\noindent\textbf{Baselines.} To demonstrate the effectiveness of our method, we compare our model with multiple baselines. We first compare with models trained from scratch with weights initialized randomly. We also compare pre-training with the classical contrastive learning method MoCo-v2~\cite{he2020momentum} with ResNet50~\cite{he2016deep} (without metadata augmentation). We also test MedAug~\cite{vu2021medaug} which uses an inter-study positive sampling strategy from the same patient.
We report our model's performance evaluated with multiple patches as our full model ($N=20$), along with an ablated version that is evaluated with $N=1$ patch per slice.
It takes $\sim$3 seconds for our full model to process one DBT slice with 20 patches.
As an ablation, we further compare with an inter-slice version of MedAug~\cite{vu2021medaug}. We experiment with two backbone models: a conventional 7-layer CNN and ResNet50~\cite{he2016deep}.

\noindent\textbf{Implementation Details.} During the pre-training stage, we train the model for 4000 epochs with a batch size of 128, a learning rate of $1.5\times 10^{-2}$, and an SGD optimizer for both backbones. The memory bank size is $S=4096$. During the fine-tuning stage, we use a learning rate of $1\times 10^{-2}$ and SGD optimizer for all three different fine-tuning strategies for 50 epochs in a discriminative fine-tuning manner. We extend the training time to 100 epochs for the baselines initialized with random weight. We chose the neighboring range $k=9$ as the range of abnormal slices. We further re-sample the slices within a batch to have a roughly 1:1 normal and abnormal ratio. We chose the model with the highest validation AUC score as the final model and reported the performance on the test set. Both the learning rate and the MoCo momentum are updated with the cosine scheduler. We chose a patch size of $448$ during fine-tuning. All the experiments are implemented with PyTorch and conducted with 4 NVIDIA A5000 GPUs. Cross entropy loss is used to optimize our model.

\noindent\textbf{Evaluation metrics.} We assess the same performance metrics for both slice- and volume-level classification. We do not report accuracy since the extreme imbalance in data distribution allows a model to produce a high accuracy by simply predicting all inputs to normal. Instead, we report the AUC, Negative Predictive Value (NPV), and Recall for each class, along with specificity at common sensitivity levels of 87\% and 80\% \cite{akselrod2019predicting}. All numbers are reported in percentages (\%).

\input{floats/fig_patch_num}

\vspace{-2mm}
\mysubsection{Slice-level Evaluation}

We first assess performance on individual slices. Slices within 9 slices of an image marked with suspicious tissue are considered abnormal, and the rest are considered normal, as the tumor/distortion is limited to a small area in the z-dimension. Thus, we have 245,875 normal and 320 abnormal slices in total. We report the results in \cref{tab:slice} and the ROC curves in \cref{fig:slice_roc}. Our model achieves the best performance for most of the metrics with a considerable gap. Even with a single patch, our model can still surpass the baselines with an AUC of $90.31\%$. When evaluated with our full model, it exceeds the baselines by 4\% of AUC and more than 10\% of specificity at both sensitivity levels. Also, we note that although the randomly initialized CNN model has the best abnormal recall, it only correctly identifies $59.10\%$ of normal slices, misclassifying nearly half of them as abnormal. Given the uneven dataset distribution, this results in almost 400 times more misclassification of normal slices.

\vspace{-2mm}
\mysubsection{Ablation Experiments}

\textbf{Fine-tuning Strategy.} We first evaluate the influence of using different fine-tuning strategies in \cref{tab:fintune} for both architectures trained with our pre-training method. We note that the discriminative fine-tuning achieved the highest performance on all three metrics for both models. The linear probing can achieve a similar level of performance as full parameter fine-tuning, which suggests our pre-trained model is able to provide robust features even without further fine-tuning.

\noindent\textbf{Number of Sampled Patches.} We further evaluate the influence of having a different number of patches during slice-level classification in \cref{fig:patch}. Increasing the number of patches can improve the performance by $\sim$4\% AUC, and most models are almost saturated when $N=4$. Considering the patch size was set to $1/4$ of the full slice size, this is reasonable.

\input{floats/tab_vol}
\input{floats/fig_vol_auc}

\mysubsection{Volume-level Evaluation}

We evaluate model performance on volume level prediction in \cref{tab:volume} and show ROC curves in \cref{fig:vol_roc}. Our full multi-patch model achieves the highest level of performance in most of the metrics. There is an increase of approximately 15\% AUC between our fine-tuned CNN model and the best baseline model. Our models achieve a specificity of 82.71\% at 87\% sensitivity and 97.81\% at 80\% sensitivity, without introducing any extra learnable parameters in volume prediction. Interestingly, we note that the volume level performance does not saturate even if the patch number is greater than 4, which suggests additional patches can help improve the model's confidence with redundant information. We also note that the high NPV for our SIFT-DBT indicates that our model has the potential to be used to filter out normal DBT scans and significantly reduce radiologists’ workload.

%% file: floats/tab_fintune.tex
\begin{table}[!t]
\centering
\resizebox{\columnwidth}{!}
{
\begin{tabular}{@{}llm{1cm}cc@{}}

\specialrule{.1em}{.05em}{.05em}
\multicolumn{1}{c}{\multirow{2}{*}{Backbone}} & \multicolumn{1}{c}{\multirow{2}{*}{\begin{tabular}[c]{@{}c@{}}Fine-tuning\\ Method\end{tabular}}} & \multicolumn{3}{c}{\textbf{BCS-DBT Slice-level Results}} \\ \cmidrule(l){3-5} 
\multicolumn{1}{c}{} & \multicolumn{1}{c}{} & AUC & Abnormal Recall & Specificity@87\% \\ \specialrule{.08em}{.05em}{.05em}
\multirow{3}{*}{CNN} & Linear Probe & 87.80 & 65.00 & 71.74 \\
 & Full Fine-tune & 87.72 & 71.56 & 68.09 \\
 & Disc. Fine-tune & \textbf{89.73} & \textbf{73.44} & \textbf{76.78} \\ \midrule
\multirow{3}{*}{ResNet50} & Linear Probe & 87.28 & 79.69 & 71.57 \\
 & Full Fine-tune & 87.94 & 75.62 & 73.43 \\
 & Disc. Fine-tune & \textbf{90.31} & \textbf{83.13} & \textbf{76.69} \\ \specialrule{.1em}{.05em}{.05em}
\end{tabular}
}
\vspace{-2mm}
\caption{\textbf{Evaluation of Different Fine-tuning Methods.} We compare the slice-level performance under a single patch setting of different fine-tuning methods following our proposed pre-training.} 
\label{tab:fintune}
\vspace{-5mm}
\end{table}

%% file: floats/fig_patch_num.tex
\begin{figure}[!t]
    \centering
    \includegraphics[width=1.0\columnwidth]{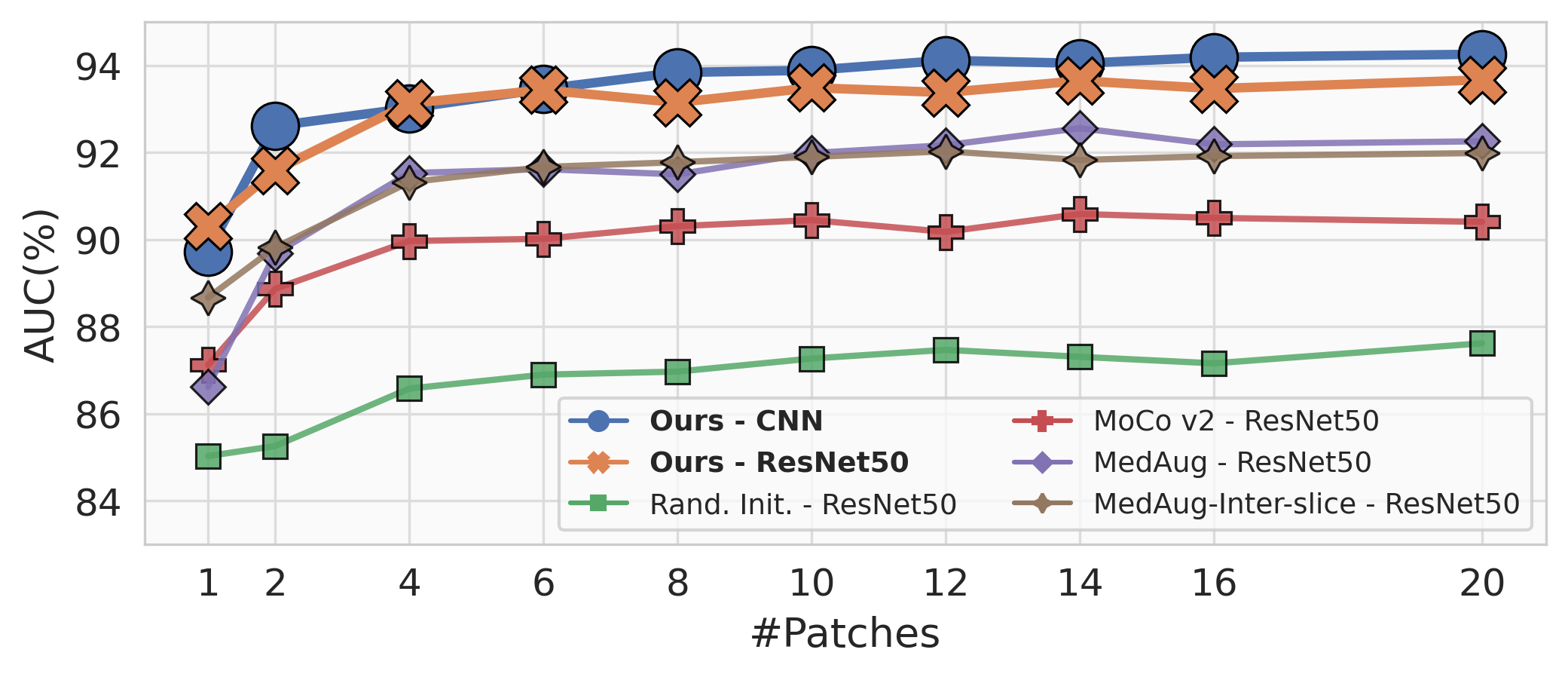}
    \vspace{-7mm}
    \caption{{\bf Evaluation with different number of patches}. We evaluate the effect of using different numbers of patches on slice classification performance (AUC) for each baseline with the ResNet50 backbone.} 
    \label{fig:patch}
    \vspace{-4mm}
\end{figure}

%% file: floats/tab_vol.tex
\begin{table}[!t]
\centering
\resizebox{\columnwidth}{!}
{
\begin{tabular}{@{}llcccccc@{}}

\specialrule{.1em}{.05em}{.05em}
\multicolumn{1}{c}{\multirow{2}{*}{Method}} & \multicolumn{1}{c}{\multirow{2}{*}{Model}} & \multicolumn{6}{c}{\textbf{BCS-DBT Volume-level Results}} \\\cmidrule(l){3-8} 
\multicolumn{1}{c}{} & \multicolumn{1}{c}{} & AUC & NPV & NR & AR & SP@87 & SP@80 \\ \specialrule{.08em}{.05em}{.05em}
\multirow{2}{*}{Rand. Init.} & CNN & 77.79 & \textbf{100.00} & 12.76 & \textbf{100.00} & 64.03 & 67.43 \\
 & ResNet50 & 65.93 & 99.38 & 64.19 & 61.54 & 21.63 & 48.31 \\ \midrule
MoCo v2 & \multirow{3}{*}{ResNet50} & 77.89 & 99.64 & 65.74 & 76.92 & 39.51 & 64.05 \\
MedAug &  & 63.95 & 99.30 & 60.47 & 58.97 & 24.87 & 35.71 \\ 
Inter-slice &  & 77.23 & 99.73 & 48.66 & 87.18 & 55.22 & 58.86 \\ \midrule
\multirow{2}{*}{Ours\textsubscript{$N$=1}} & CNN & 78.85 & 99.64 & 74.73 & 74.35 & 52.84 & 65.85 \\
  & ResNet50 & 69.55 & 99.67 & 24.36 & {\ul 92.31} & 31.05 & 55.89 \\ \hdashline
\multirow{2}{*}{Ours} & CNN & {\ul 92.58} & {\ul 99.81} & \textbf{84.15} & 84.62 & \textbf{82.71} & {\ul 93.52} \\
 & ResNet50 & \textbf{92.69} & {\ul 99.81} & {\ul 83.97} & 84.62 & {\ul 80.86} & \textbf{97.81} \\ \specialrule{.1em}{.05em}{.05em}
\end{tabular}
}
\vspace{-2mm}
\caption{\textbf{Volume-level Classification Results.}
We highlight the top result in bold and the second-best with an underline. NR, Normal Recall; AR, Abnormal Recall; SP@\#, Specificity at \#\% sensitivity.} 
\label{tab:volume}
\vspace{-3mm}
\end{table}

%% file: floats/fig_vol_auc.tex
\begin{figure}[!t]
    \centering
    \includegraphics[width=1.0\columnwidth]{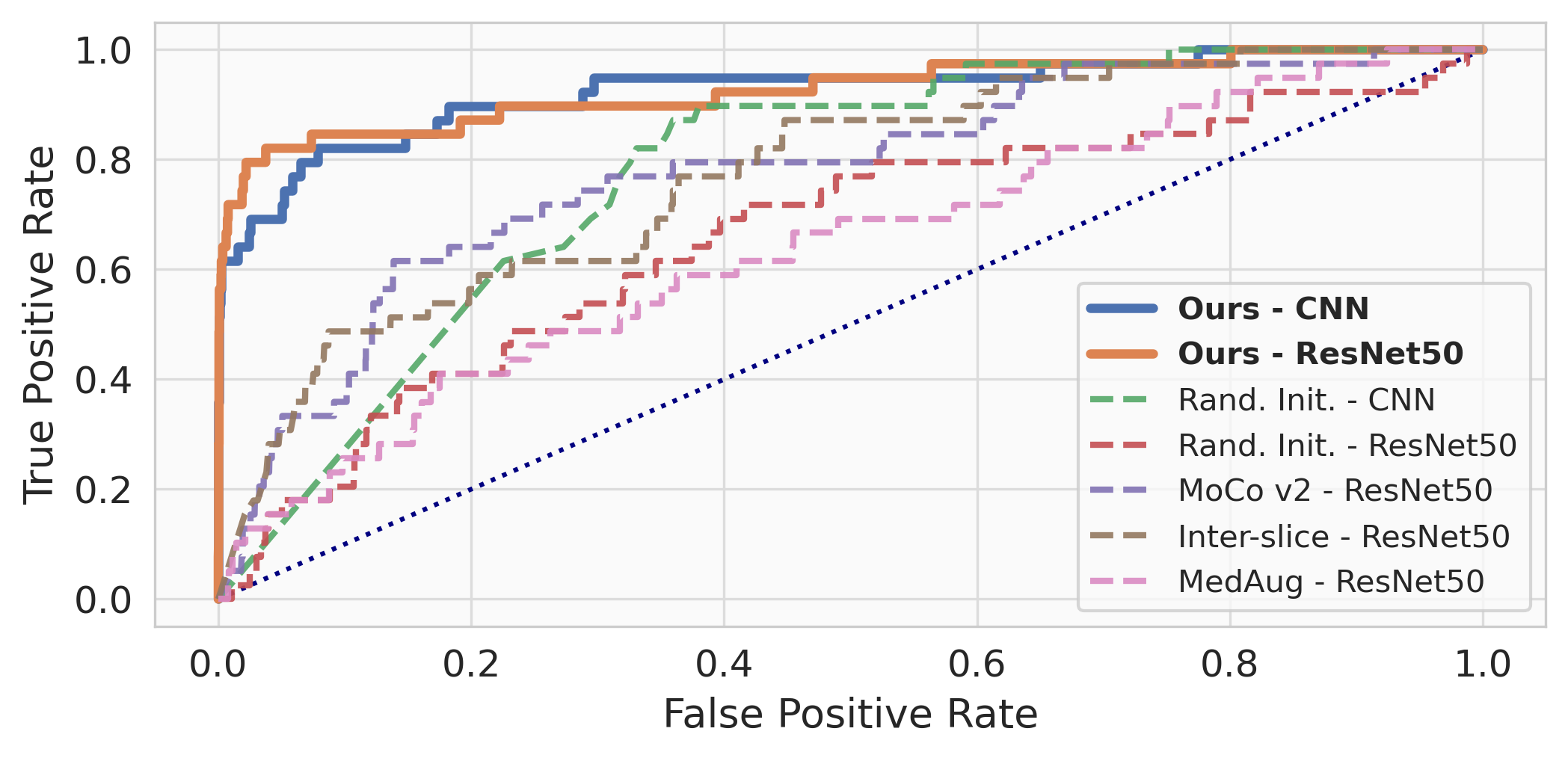}
    \vspace{-7mm}
    \caption{{\bf Volume-level ROC curve}. We plot the ROC curve for the volume-level evaluation. Our method is plotted with a solid line.}
    \label{fig:vol_roc}
    \vspace{-4mm}
\end{figure}

%% file: parts/4_conclusion.tex
\mysection{Discussion and Conclusion}

We proposed a novel slice- and volume-level classification framework, SIFT-DBT, with both self-supervised pre-trained weight initialization and fine-tuning to address the extremely imbalanced data distribution in real-world DBT data. We demonstrate the effectiveness of our method via extensive experiments and detailed evaluation. Our method surpasses all the baselines with a considerable gap on multiple metrics and may potentially improve the workflow of radiologists, allowing them to concentrate more on high-risk scans.

%% file: parts/5_acknowledge.tex
\mysection{Compliance with Ethical Standards}

This research study was conducted retrospectively using human subject data made available in open access on The Cancer Imaging Archive \cite{clark2013cancer} by the Mazurowski lab at Duke University. Ethical approval was not required as confirmed by the license attached with the open-access data.

\mysection{Acknowledgements}

This work was supported by NIH grant R21EB032950. We thank Jiazhen Zhang for the helpful discussions.

%% file: main.bbl
\begin{thebibliography}{10}

\bibitem{GLOBOCAN}
H.~Sung, J.~Ferlay, R.L. Siegel, M.~Laversanne, I.~Soerjomataram, A.~Jemal, and F.~Bray,
\newblock ``Global cancer statistics 2020: Globocan estimates of incidence and mortality worldwide for 36 cancers in 185 countries,''
\newblock {\em CA: A Cancer Journal for Clinicians}, vol. 71, no. 3, pp. 209--249, 2021.

\bibitem{hooley2017advances}
R.J. Hooley, M.A. Durand, and L.E. Philpotts,
\newblock ``Advances in digital breast tomosynthesis,''
\newblock {\em American Journal of Roentgenology}, vol. 208, no. 2, pp. 256--266, 2017.

\bibitem{konz2023competition}
N.~Konz, M.~Buda, H.~Gu, A.~Saha, J.~Yang, J.~Chlkedowski, J.~Park, J.~Witowski, J.J. Geras, Y.~Shoshan, et~al.,
\newblock ``A competition, benchmark, code, and data for using artificial intelligence to detect lesions in digital breast tomosynthesis,''
\newblock {\em JAMA Network Open}, vol. 6, no. 2, pp. e230524--e230524, 2023.

\bibitem{buda2020detection}
M.~Buda, A.~Saha, R.~Walsh, S.~Ghate, N.~Li, A.~'Swikecicki, J.Y. Lo, and M.A. Mazurowski,
\newblock ``Detection of masses and architectural distortions in digital breast tomosynthesis: a publicly available dataset of 5,060 patients and a deep learning model,''
\newblock {\em arXiv preprint arXiv:2011.07995}, 2020.

\bibitem{zhang2023deep}
Y.~Zhang, B.~Kang, B.~Hooi, S.~Yan, and J.~Feng,
\newblock ``Deep long-tailed learning: A survey,''
\newblock {\em IEEE Transactions on Pattern Analysis and Machine Intelligence}, 2023.

\bibitem{li2023hierarchical}
Y.~Li, G.~Qian, X.~Jiang, Z.~Jiang, W.~Wen, S.~Zhang, K.~Li, and Q.~Lao,
\newblock ``Hierarchical-instance contrastive learning for minority detection on imbalanced medical datasets,''
\newblock {\em IEEE Transactions on Medical Imaging}, 2023.

\bibitem{lin2017focal}
T.Y. Lin, P.~Goyal, R.~Girshick, K.~He, and P.~Doll{\'a}r,
\newblock ``Focal loss for dense object detection,''
\newblock in {\em Proceedings of the IEEE international conference on computer vision}, 2017, pp. 2980--2988.

\bibitem{lee2023transformer}
W.~Lee, H.~Lee, H.~Lee, E.K. Park, H.~Nam, and T.~Kooi,
\newblock ``Transformer-based deep neural network for breast cancer classification on digital breast tomosynthesis images,''
\newblock {\em Radiology: Artificial Intelligence}, vol. 5, no. 3, pp. e220159, 2023.

\bibitem{tardy2021trainable}
M.~Tardy and D.~Mateus,
\newblock ``Trainable summarization to improve breast tomosynthesis classification,''
\newblock in {\em International Conference on Medical Image Computing and Computer-Assisted Intervention}. Springer, 2021, pp. 140--149.

\bibitem{oord2018representation}
A.V.D. Oord, Y.~Li, and O.~Vinyals,
\newblock ``Representation learning with contrastive predictive coding,''
\newblock {\em arXiv preprint arXiv:1807.03748}, 2018.

\bibitem{he2020momentum}
K.~He, H.~Fan, Y.~Wu, S.~Xie, and R.~Girshick,
\newblock ``Momentum contrast for unsupervised visual representation learning,''
\newblock in {\em Proceedings of the IEEE/CVF conference on computer vision and pattern recognition}, 2020, pp. 9729--9738.

\bibitem{vu2021medaug}
Y.N.T. Vu, R.~Wang, N.~Balachandar, C.~Liu, A.Y. Ng, and P.~Rajpurkar,
\newblock ``Medaug: Contrastive learning leveraging patient metadata improves representations for chest x-ray interpretation,''
\newblock in {\em Machine Learning for Healthcare Conference}. PMLR, 2021, pp. 755--769.

\bibitem{howard2018universal}
J.~Howard and S.~Ruder,
\newblock ``Universal language model fine-tuning for text classification,''
\newblock {\em arXiv preprint arXiv:1801.06146}, 2018.

\bibitem{he2016deep}
K.~He, X.~Zhang, S.~Ren, and J.~Sun,
\newblock ``Deep residual learning for image recognition,''
\newblock in {\em Proceedings of the IEEE conference on computer vision and pattern recognition}, 2016, pp. 770--778.

\bibitem{akselrod2019predicting}
A.~Akselrod-Ballin, M.~Chorev, Y.~Shoshan, A.~Spiro, A.~Hazan, R.~Melamed, E.~Barkan, E.~Herzel, S.~Naor, E.~Karavani, et~al.,
\newblock ``Predicting breast cancer by applying deep learning to linked health records and mammograms,''
\newblock {\em Radiology}, vol. 292, no. 2, pp. 331--342, 2019.

\bibitem{clark2013cancer}
K.~Clark, B.~Vendt, K.~Smith, J.~Freymann, J.~Kirby, P.~Koppel, S.~Moore, S.~Phillips, D.~Maffitt, M.~Pringle, et~al.,
\newblock ``The cancer imaging archive (tcia): maintaining and operating a public information repository,''
\newblock {\em Journal of digital imaging}, vol. 26, pp. 1045--1057, 2013.

\end{thebibliography}
